\documentclass{vak2024}
\usepackage[utf8]{inputenc}
\usepackage{url}
\usepackage[pdfpagelabels=false]{hyperref}	% Hyperlinks
\hypersetup{colorlinks=true,linkcolor=blue,citecolor=blue,filecolor=blue,urlcolor=blue,}

\usepackage{aas_macros}

\newcommand{\pcmm}{~cm$^{-3}$}	% per cm-cubed
\newcommand{\kms}{~km\,s$^{-1}$}
\newcommand{\Hii}{H~{\sc ii}}

\newcommand{\Feii}{Fe~{\sc ii}}

\begin{document}
\title{Study of the high-mass star-forming region S255IR at various scales}
\titlerunning{S255IR at various scales}  % abbreviated title (for running head)
	%also used for the TOC unless
	%\toctitle is used
\author{I.~I.~Zinchenko\inst{1}\and S.-Y.~Liu\inst{2}\and D.~K.~Ojha\inst{3}\and Y.-N.~Su\inst{2}\and P.~M. Zemlyanukha\inst{1}
}
\authorrunning{Zinchenko et al.} % abbreviated author list (for running head)
    %\tocauthor{}
	%
\institute{Federal Research Center A.V. Gaponov-Grekhov Institute of Applied Physics of the Russian Academy of Sciences, 46 Ul’yanov str., Nizhny Novgorod 603950, Russia
	    \and Institute of Astronomy and Astrophysics, Academia Sinica, 11F of ASMAB, AS/NTU No.1, Sec. 4, Roosevelt Rd, Taipei 10617, Taiwan
     \and Department of Astronomy and Astrophysics, Tata Institute of Fundamental Research, Homi Bhabha Road, Mumbai 400005, India
}

\abstract{
The S255IR-SMA1 core contains the protostar NIRS3 with a mass of $\sim$20~M$_\odot$. Several years ago, the first burst of luminosity for massive protostars, caused by an episodic accretion event, was recorded here. We have been studying this object for a long time using various instruments, including ALMA. The general morphology and kinematics of this area have been investigated. Disk-shaped structures, jets and outflows have been identified and studied in detail. We recently observed this object with ALMA with a resolution an order of magnitude higher than previously achieved -- about 15 milliarcseconds, which corresponds to about 25 AU. This paper presents new results from the analysis of these data together with observations in other bands. The new data show an inhomogeneous disk structure, an ionized region around the protostar, and the presence of a jet observed in the submillimeter continuum, consisting of individual knots, the orientation of which differs markedly from that on large scales. The submillimeter emission from the jet most likely represents bremsstrahlung from ionized gas. Based on observations of the lines of some molecules, the kinematics and physical characteristics of this region are discussed. Methanol maser emission associated with the jet is observed.

\doi{10.26119/VAK2024-ZZZZ}
}

\maketitle
\section{Introduction}
There are still several competing scenarios of high mass star formation \citep[e.g.,][]{Tan14, Motte18, Padoan2020, Rosen2020}. A key question is whether this process is a scaled-up version of the low-mass star formation or is significantly different. Observations at various scales are essential for selection between different scenarios. 

In recent years, a large attention has been paid to the luminosity bursts from massive protostars, which are believed to be caused by episodic disk-mediated accretion events. There are theoretical models which predict such a behavior \citep[e.g.,][]{Meyer17, Meyer19}. To date, several such bursts have been recorded \citep[see the summary in][]{Wolf2024}. One of the first of them was the burst in S255IR NIRS3, which was observed at IR \citep{Caratti16} and submillimeter \citep{Liu18} wavelengths, and was accompanied by the methanol maser flare \citep{Moscadelli17, Szymczak2018}.

The large star-forming complex sandwiched between the evolved \Hii\ regions S255 and S257 \citep{Ojha2011} is a well-known and actively investigated area of high-mass star formation. It contains two major star-forming sites: S255IR and S255N. %(Figure~\ref{fig:large}). 
Here we focus on the first one. %A detailed investigation of S255N was presented by \citet{Zemlyanukha2018}.

%\begin{figure}[htb]
%    \centering
%    \includegraphics[width=0.7\textwidth]{figs/s255_1mm+610+1280}
%    \caption{A composite image of the S255--S257 area. The red color shows the 1.2~mm continuum emission \citep{Zin09}, green and blue colors show the 1280 and 610~MHz GMRT images, respectively \citep{Zin12}.}
%    \label{fig:large}
%\end{figure}

The distance to S255IR is estimated at $1.78_{-0.11}^{+0.12}$~kpc from the maser parallax measurements \citep{Burns16}. It contains three major cores SMA1, SMA2 and SMA3 \citep{Wang11} and several smaller condensations \citep{Zin2020}. The SMA1 core harbors a $\sim$20~M$_\odot$ protostar NIRS3 \citep{Zin15}. The mass is estimated from the bolometric luminosity of $\sim 3\times 10^4$~L$_\odot$ at the adopted distance.
Here we summarize and discuss the main results of our investigations of this object.

\section{Observational data} \label{sec:obs}
We have observed the S255IR area with several single-dish radio telescopes (IRAM-30m, OSO-29m, MPIfR 100-m radiotelescope in Effelsberg) and with radio interferometers (ALMA, GMRT, SMA, VLA). These observations include imaging in continuum and in many molecular lines. The frequency coverage was from $\sim$600~MHz (GMRT) to $\sim$350~GHz (ALMA). The angular resolution was from $\sim$40 arcseconds (for single-dish observations) to $\sim$15 milliarcseconds (for ALMA observations). At the distance to S255IR 1 arcsecond corresponds to $\sim$1800~au.
In addition, we combine our radio data with our and other available data in other bands, in particular with IR observations. 

\section{Structure of S255IR at various scales} \label{sec:structure}

\begin{figure}[htb]
    \centering
    \includegraphics[width=0.8\textwidth]{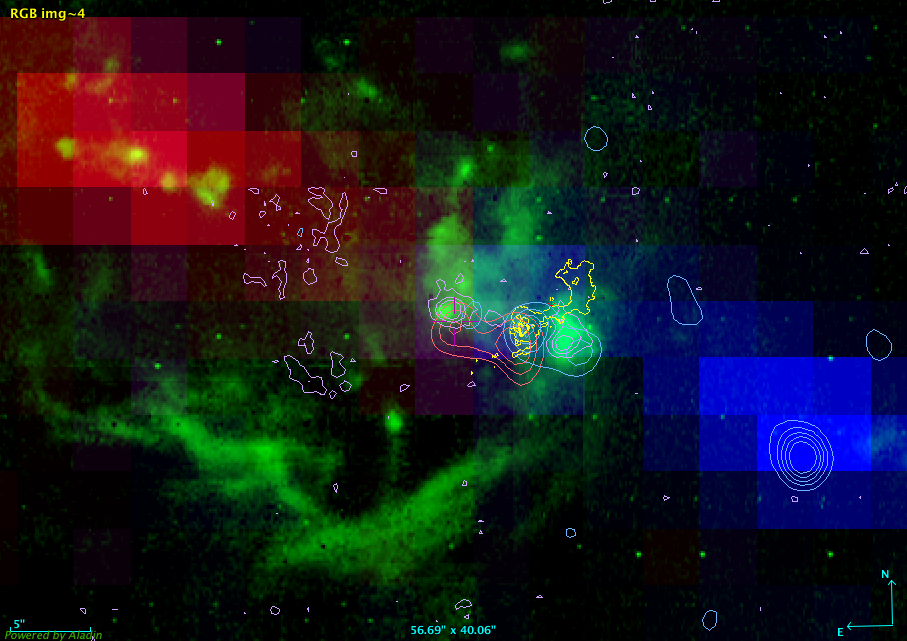}
    \caption{S255IR area in the 2.12 $\mu$m H$_2$ line (green) and high-velocity CO(3--2) emission (red and blue). The yellow contours show the 0.9 mm emission (ALMA), the orange and blue contours show the high-velocity HCN(4--3) emission, the magenta contours show the 1.64 $\mu$m \Feii\ emission. The NIR data are from SINFONI observations \citep{Wang11}. The radio data are from \citet{Zin12, Zin15, Zin2020}.
}
    \label{fig:h2}
\end{figure}

In Figure~\ref{fig:h2} we present a composite image of the S255IR area (size $\sim$1$^\prime$ or $\sim$0.5~pc) in several lines and in continuum.
%in the 2.12 $\mu$m H$_2$ line (green) and high-velocity CO(3--2) emission (red and blue). The yellow contours show the 0.9 mm continuum emission (ALMA), the orange and blue contours show the high-velocity HCN(4--3) emission, the magenta contours show the 1.64 $\mu$m \Feii\ emission. The NIR data are from SINFONI observations \citep{Wang11}. The radio data are from \citet{Zin12, Zin15, Zin2020}. 

There are several H$_2$ knots aligned along the IR jet direction (PA $\approx$ 67$^\circ$) observed many years ago \citep{Howard97}. The knots closest to S255IR-SMA1 show also a strong \Feii\ emission and are associated with the high-velocity HCN(4--3) and HCO$^+$(4--3) emission \citep{Zin15}. These knots also coincide with the radio knots with non-thermal radio spectra \citep{Obonyo2021}. Apparently, these knots are associated with bow shocks and dense molecular gas. With the projected expansion speed of $\sim$450\kms\ for the NE lobe \citep{Fedriani2023, Cesaroni2023}  and $\sim$285\kms\ for the SW lobe \citep{Cesaroni2024} their ejection happened about 60--70 years ago. Assuming the same velocity, the most distant knots were ejected several hundred years ago.
%\citet{Obonyo2021} estimated that the NE and SW knots were ejected from the central source about 78 and 39 years ago, respectively, assuming the jet velocity of 500~\kms. Later, the projected expansion speed was estimated as $\sim$450\kms\ for the NE lobe \citep{Fedriani2023, Cesaroni2023}  and $\sim$285\kms\ for the SW lobe \citep{Cesaroni2024}. In this case the ejection happened about 60--70 years ago. 

The far H$_2$ NE knots coincide with the red-shifted CO(3--2) outflow lobe observed with the IRAM-30m radio telescope (Figure~\ref{fig:h2}). The H$_2$ knots in SW direction are less pronounced (although visible). The peak of the SW CO(3--2) outflow lobe coincides with the dense high-velocity clump seen in the HCN(4--3) and CS(7--6) lines \citep{Zin15}. It is worth noting that there are two almost parallel outflows here, originating at the SMA1 and SMA2 cores \citep{Zin2020}. The associations mentioned above indicate that the CO(3--2) outflow observed with the IRAM-30m radio telescope originates at the SMA1 core.

\begin{figure}[htb]
    \centering
    \includegraphics[width=0.30\textwidth]{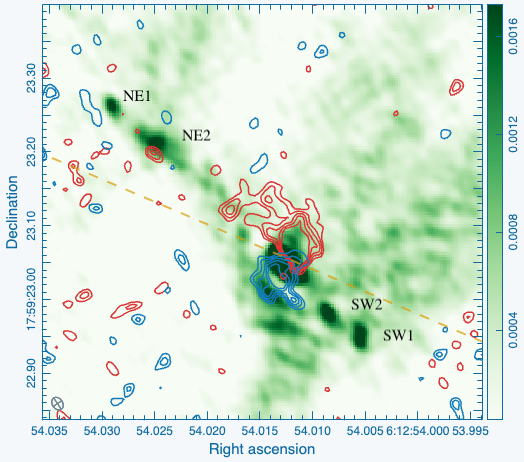}
    \hfill
%    \includegraphics[width=0.34\textwidth]{figs/c34s_mom1}
%    \hfill
    \includegraphics[width=0.29\textwidth]{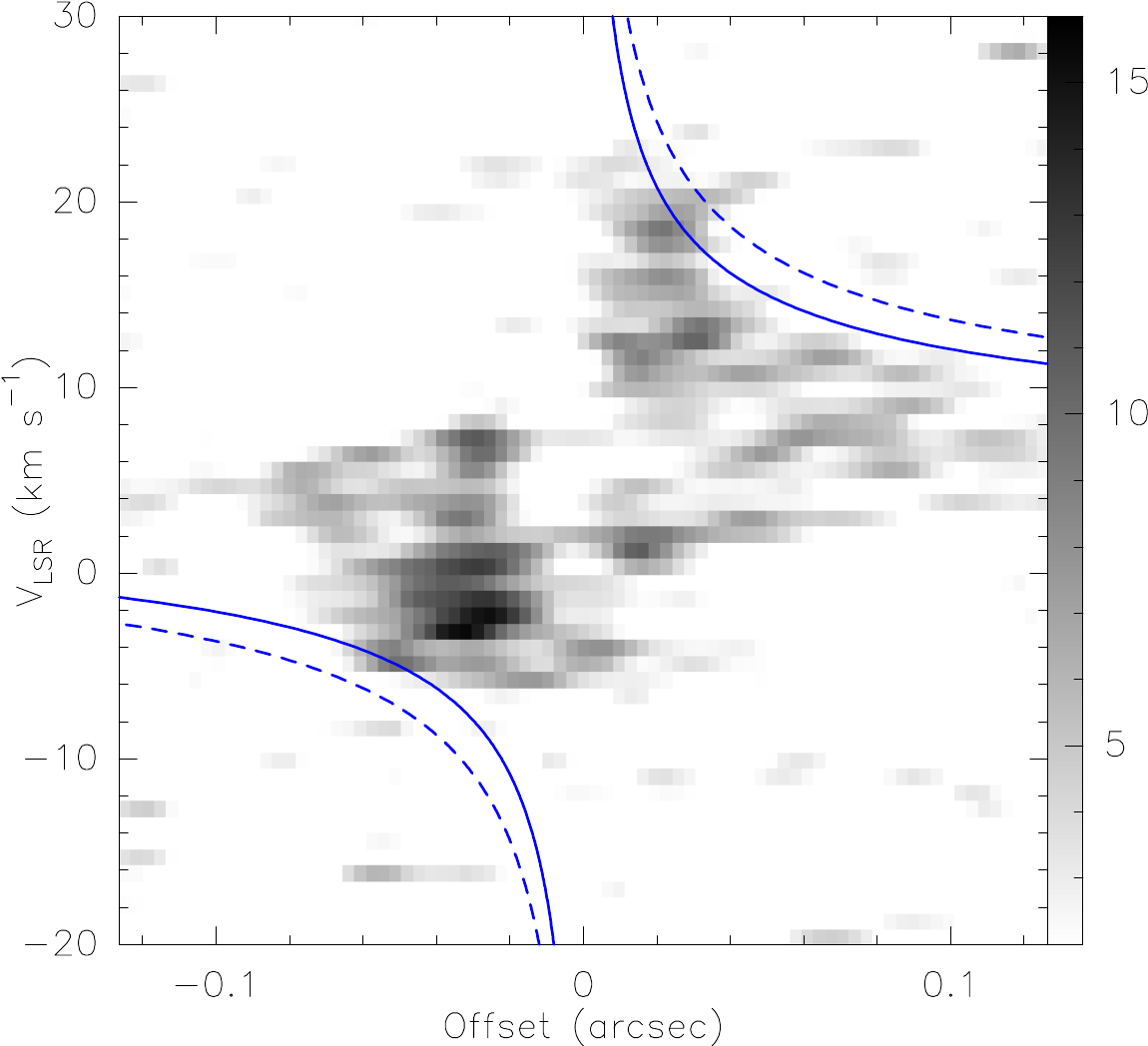}
    \hfill
    \includegraphics[width=0.39\textwidth]{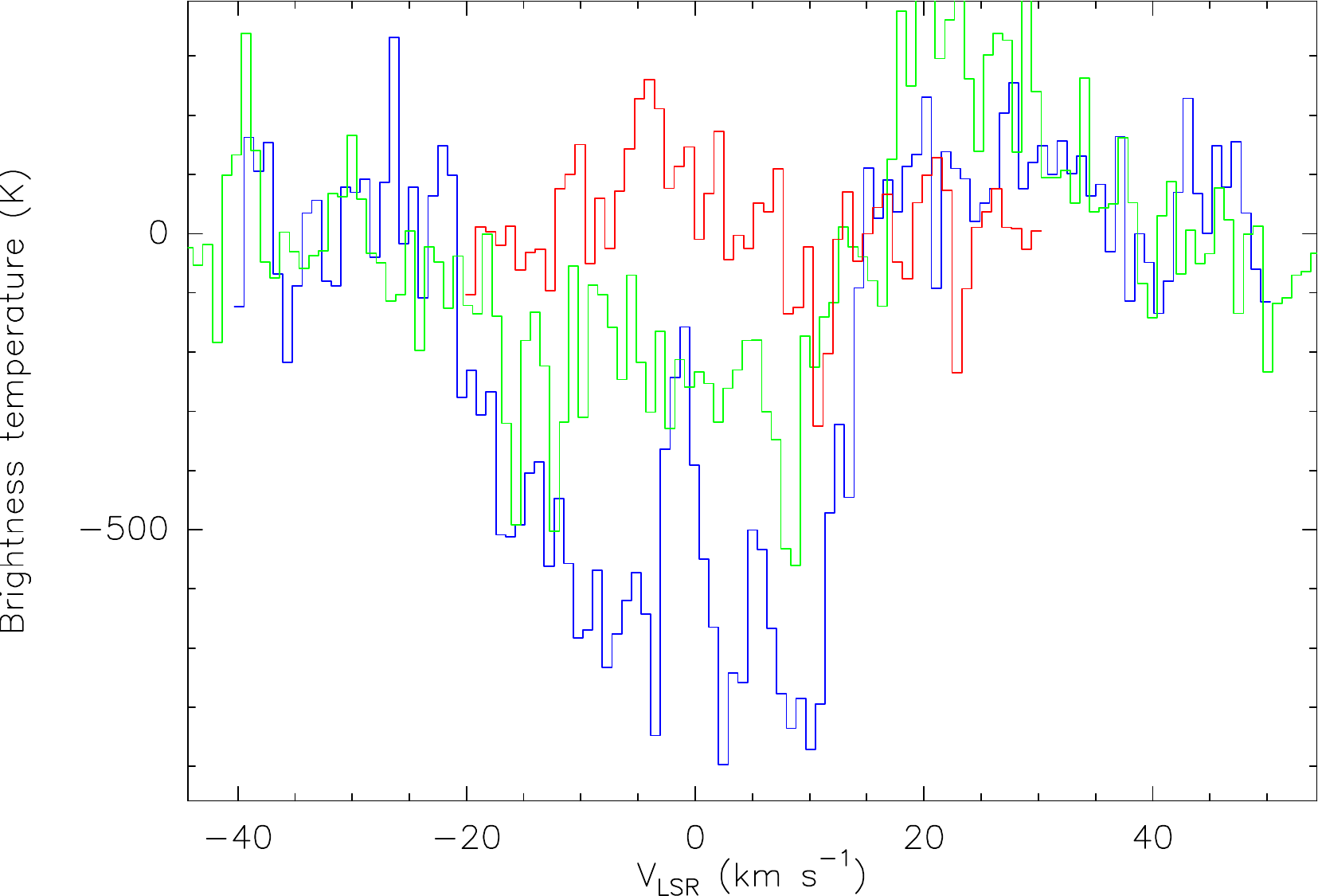}
    \caption{Left panel: the image of S255IR area in continuum obtained with ALMA at 0.9 mm with $\sim$15~mas resolution (Zinchenko, Liu \& Su, in preparation). The knots in the jet are marked. The red and blue contours show the emission in the C$^{34}$S(7--6) line wings. The dashed line indicates the jet orientation at larger scales (PA = 67$^\circ$). 
    %Central panel: the 1$^{st}$ moment map of the S255IR area in the C$^{34}$S(7--6) line overlaid with contours of the 0$^{th}$ moment of the C$^{34}$S(7--6) line emission (thick black lines) and contours of the 0.9~mm continuum emission (thin green lines). 
    Central panel: the PV diagram in the C$^{34}$S(7--6) line along the disk major axis. The curves correspond to Keplerian rotation around the central mass of $M \sin^2i = 10$~M$_\odot$ (solid) and $M \sin^2i = 15$~M$_\odot$ (dashed), where $i$ is the inclination angle.
    Right panel: spectra of the C$^{34}$S(7--6) (red), SiO(8--7) and CO(3--2) (blue) emission toward the central continuum peak.}
    \label{fig:jet}
\end{figure}

Our ALMA observations at $\sim$150~mas resolution show that this core represents a rotating and infalling envelope (pseudo-disk) around the NIRS3 protostar \citep{Liu2020}. In Figure~\ref{fig:jet} we present some results of our latest (performed in September 2021) ALMA observations of this object with an order of magnitude higher resolution of $\sim$15~mas, which corresponds to $\sim$25~au (Zinchenko, Liu \& Su, A\&A, submitted).

The continuum image shows the central bright source (the brightness temperature is $\sim$850~K), which practically coincides with the NIRS3 position, and two pairs of bright knots ($\sim$80--110~K), one pair in each outflow lobe, located almost on a strait line along with the central source. The position angle of this line is approximately 47$^\circ$. Apparently, these knots belong to the jet emanating from the central source. It is worth noting that the central source is elongated approximately in the direction of the jet. The position angle of the jet differs by $\approx 20^\circ$ from that observed at larger scales (Figure~\ref{fig:jet}), as found also in some other observations at small scales \citep{Hirota2021, Cesaroni2023}. These results indicate the jet precession as suggested in some previous works \citep{Obonyo2021, Cesaroni2023}. 
The pairs of knots imply two ejection events with the time interval about 1.5~yr. This agrees well with the 6.7~GHz methanol maser light curve \citep{Szymczak2018}. 

The high brightness of the central source and its morphology imply a significant contribution of the free-free emission. Taking into account the flux measurements at lower frequencies from $\sim$3 to $\sim$92~GHz \citep{Obonyo2021, Cesaroni2023, Cesaroni2024} we estimated contributions of the ionized gas and dust emission assuming an optically thin free-free component. For the ionized gas we obtain the emission measure of $EM\sim  10^{10}$~pc\,cm$^{-6}$ and the electron density of $n_\mathrm{e}\sim 10^7$\pcmm. Such properties are typical for hypercompact \Hii\ regions. It is probably surrounded by a dust cocoon. An alternative model assumes a partly optically thick hypercompact \Hii\ region, which explains better the spectral index in the range 92--340~GHz.
In the emission of the knots in the jet, the free-free component apparently dominates.

Figure~\ref{fig:jet} shows a disk-like rotating structure around the central source. The molecular emission is very inhomogeneous which implies a clumpiness. The PV diagrams in several lines (see an example in Figure~\ref{fig:jet}) indicate a sub-Keplerian rotation. There are also deep absorption features in the molecular spectra toward the bright central source. The deepest features are red-shifted relative the systemic velocity of the core, which implies an infall.

\begin{figure}[htb]
    \centering
    \includegraphics[width=0.38\textwidth]{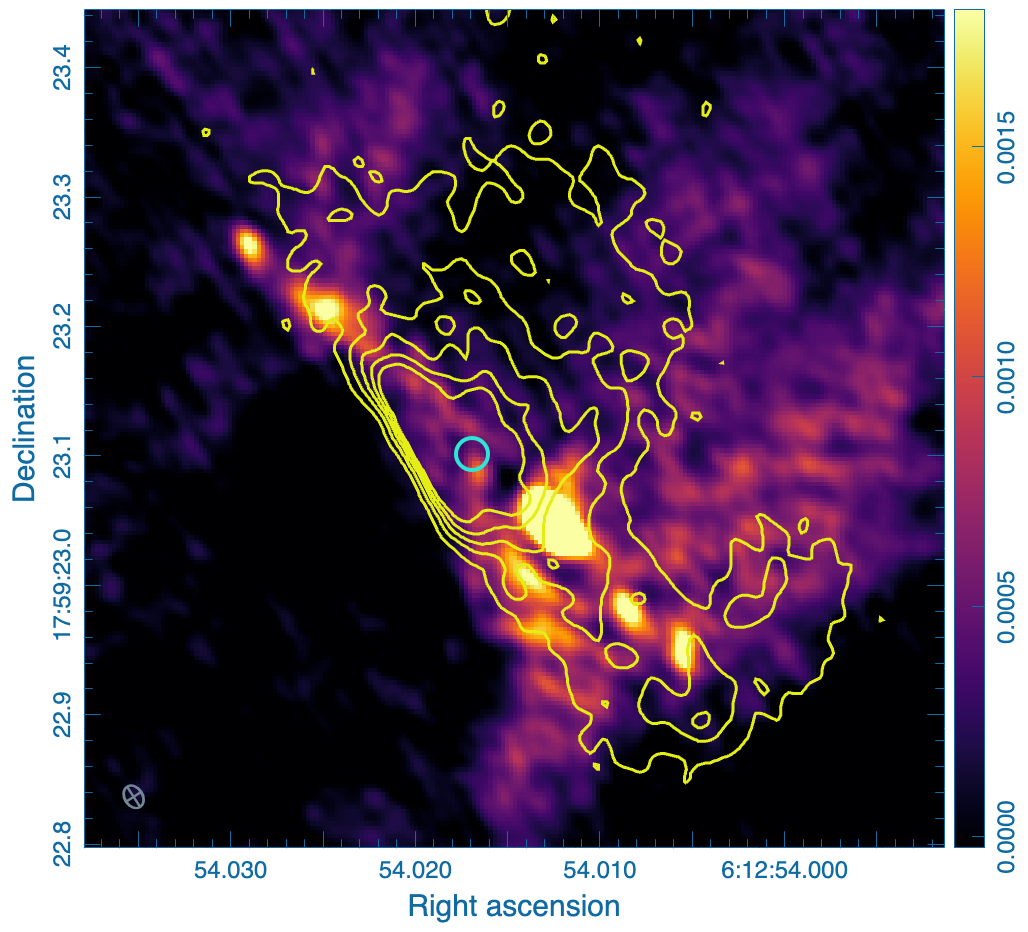}
    \hspace{2mm}
    \includegraphics[width=0.5\textwidth]{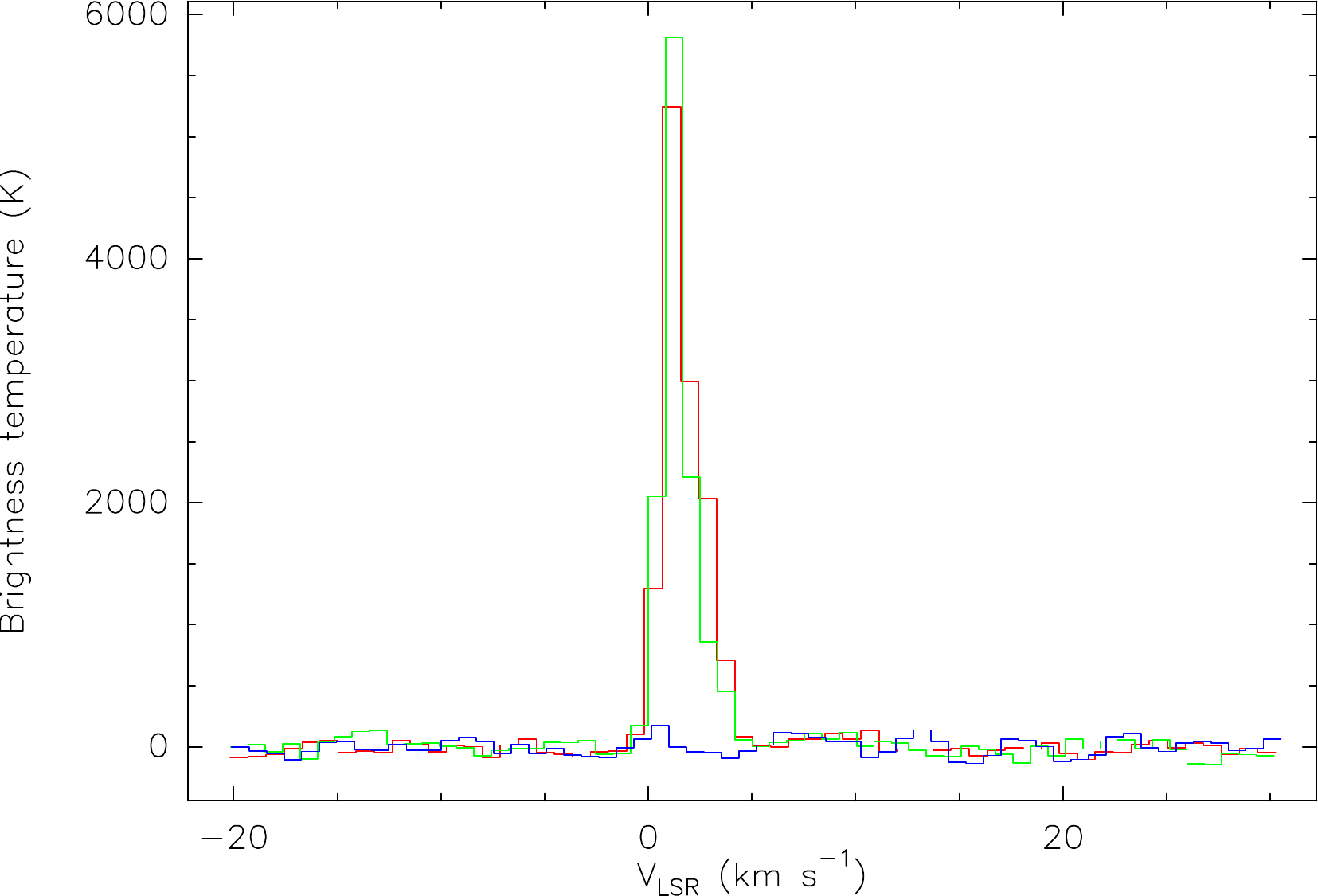}
    \caption{Left panel: the continuum image of S255IR at 0.9~mm overlaid with contours of the CH$_3$OH $12_1 - 12_0$ A$^{-+}$ emission. Right panel: spectra of the CH$_3$OH $12_1 - 12_0$ A$^{-+}$ (red), $14_1 - 14_0$ A$^{-+}$ (green) and $^{13}$CH$_3$OH $14_1 - 14_0$ A$^{-+}$ (blue) emission at the position marked by the circle.}
    \label{fig:maser}
\end{figure}

Earlier we detected a new methanol maser line ($14_1 - 14_0$~A$^{-+}$ at 349.1~GHz) toward S255IR-SMA1 \citep{Zin17}. In the new data its intensity is consistent with the previous measurements, which show a decay since 2016 \citep{Salii2022}. Now, there was one more line of this series in the observed bands, $12_1 - 12_0$~A$^{-+}$. It also shows the maser effect (Figure~\ref{fig:maser}). These masers are apparently associated with the jet and coincide with some other methanol masers discovered here recently \citep{Baek2023}.
%The maser effect in the lines of this series can be a tracer of luminosity flares during high-mass star formation \citep{Salii2022}.

%\section{Discussion} \label{sec:disc}

\section{Conclusions}
In general, observations of S255IR confirm the scenario of disk accretion as a formation mechanism for $\sim$20~M$_\odot$ stars. Further studies of its evolution after the recent accretion burst would be very important.

%\acknowledgements{This research has made use of the SIMBAD database, CDS, Strasbourg Astronomical Observatory, France.}

\section*{Funding} 
This work was supported by the Russian Science Foundation grant No 24-12-00153.

\bibliographystyle{aa}
\bibliography{s255ir}

\begin{thebibliography}{27}
\expandafter\ifx\csname natexlab\endcsname\relax\def\natexlab#1{#1}\fi

\bibitem[{{Baek} {et~al.}(2023){Baek}, {Lee}, {Evans}, {Hirota}, {Aikawa}, {Kang}, {Kim}, \& {J{\o}rgensen}}]{Baek2023}
{Baek}, G., {Lee}, J.-E., {Evans}, N.~J., {et~al.} 2023, \apjl, 954, L25

\bibitem[{{Burns} {et~al.}(2016){Burns}, {Handa}, {Nagayama}, {Sunada}, \& {Omodaka}}]{Burns16}
{Burns}, R.~A., {Handa}, T., {Nagayama}, T., {Sunada}, K., \& {Omodaka}, T. 2016, \mnras, 460, 283

\bibitem[{{Caratti O Garatti} {et~al.}(2017){Caratti O Garatti}, {Stecklum}, {Garcia Lopez}, {Eisl{\"o}ffel}, {Ray}, {Sanna}, {Cesaroni}, {Walmsley}, {Oudmaijer}, {de Wit}, {Moscadelli}, {Greiner}, {Krabbe}, {Fischer}, {Klein}, \& {Iba{\~n}ez}}]{Caratti16}
{Caratti O Garatti}, A., {Stecklum}, B., {Garcia Lopez}, R., {et~al.} 2017, Nature Physics, 13, 276

\bibitem[{{Cesaroni} {et~al.}(2023){Cesaroni}, {Moscadelli}, {Caratti o Garatti}, {Eisl{\"o}ffel}, {Fedriani}, {Neri}, {Ray}, {Sanna}, \& {Stecklum}}]{Cesaroni2023}
{Cesaroni}, R., {Moscadelli}, L., {Caratti o Garatti}, A., {et~al.} 2023, \aap, 680, A110

\bibitem[{{Cesaroni} {et~al.}(2024){Cesaroni}, {Moscadelli}, {Caratti o Garatti}, {Eisl{\"o}ffel}, {Fedriani}, {Neri}, {Ray}, {Sanna}, \& {Stecklum}}]{Cesaroni2024}
{Cesaroni}, R., {Moscadelli}, L., {Caratti o Garatti}, A., {et~al.} 2024, \aap, 683, L15

\bibitem[{{Fedriani} {et~al.}(2023){Fedriani}, {Caratti o Garatti}, {Cesaroni}, {Tan}, {Stecklum}, {Moscadelli}, {Koutoulaki}, {Cosentino}, \& {Whittle}}]{Fedriani2023}
{Fedriani}, R., {Caratti o Garatti}, A., {Cesaroni}, R., {et~al.} 2023, \aap, 676, A107

\bibitem[{{Hirota} {et~al.}(2021){Hirota}, {Cesaroni}, {Moscadelli}, {Sugiyama}, {Burns}, {Kim}, {Sunada}, \& {Yonekura}}]{Hirota2021}
{Hirota}, T., {Cesaroni}, R., {Moscadelli}, L., {et~al.} 2021, \aap, 647, A23

\bibitem[{{Howard} {et~al.}(1997){Howard}, {Pipher}, \& {Forrest}}]{Howard97}
{Howard}, E.~M., {Pipher}, J.~L., \& {Forrest}, W.~J. 1997, \apj, 481, 327

\bibitem[{{Liu} {et~al.}(2020){Liu}, {Su}, {Zinchenko}, {Wang}, {Meyer}, {Wang}, \& {Hsieh}}]{Liu2020}
{Liu}, S.-Y., {Su}, Y.-N., {Zinchenko}, I., {et~al.} 2020, \apj, 904, 181

\bibitem[{{Liu} {et~al.}(2018){Liu}, {Su}, {Zinchenko}, {Wang}, \& {Wang}}]{Liu18}
{Liu}, S.-Y., {Su}, Y.-N., {Zinchenko}, I., {Wang}, K.-S., \& {Wang}, Y. 2018, \apj, 863, L12

\bibitem[{{Meyer} {et~al.}(2019){Meyer}, {Vorobyov}, {Elbakyan}, {Stecklum}, {Eisl{\"o}ffel}, \& {Sobolev}}]{Meyer19}
{Meyer}, D.~M.~A., {Vorobyov}, E.~I., {Elbakyan}, V.~G., {et~al.} 2019, \mnras, 482, 5459

\bibitem[{{Meyer} {et~al.}(2017){Meyer}, {Vorobyov}, {Kuiper}, \& {Kley}}]{Meyer17}
{Meyer}, D.~M.-A., {Vorobyov}, E.~I., {Kuiper}, R., \& {Kley}, W. 2017, \mnras, 464, L90

\bibitem[{{Moscadelli} {et~al.}(2017){Moscadelli}, {Sanna}, {Goddi}, {Walmsley}, {Cesaroni}, {Caratti o Garatti}, {Stecklum}, {Menten}, \& {Kraus}}]{Moscadelli17}
{Moscadelli}, L., {Sanna}, A., {Goddi}, C., {et~al.} 2017, \aap, 600, L8

\bibitem[{{Motte} {et~al.}(2018){Motte}, {Bontemps}, \& {Louvet}}]{Motte18}
{Motte}, F., {Bontemps}, S., \& {Louvet}, F. 2018, \araa, 56, 41

\bibitem[{{Obonyo} {et~al.}(2021){Obonyo}, {Lumsden}, {Hoare}, {Kurtz}, \& {Purser}}]{Obonyo2021}
{Obonyo}, W.~O., {Lumsden}, S.~L., {Hoare}, M.~G., {Kurtz}, S.~E., \& {Purser}, S.~J.~D. 2021, \mnras, 501, 5197

\bibitem[{{Ojha} {et~al.}(2011){Ojha}, {Samal}, {Pandey}, {Bhatt}, {Ghosh}, {Sharma}, {Tamura}, {Mohan}, \& {Zinchenko}}]{Ojha2011}
{Ojha}, D.~K., {Samal}, M.~R., {Pandey}, A.~K., {et~al.} 2011, \apj, 738, 156

\bibitem[{{Padoan} {et~al.}(2020){Padoan}, {Pan}, {Juvela}, {Haugb{\o}lle}, \& {Nordlund}}]{Padoan2020}
{Padoan}, P., {Pan}, L., {Juvela}, M., {Haugb{\o}lle}, T., \& {Nordlund}, {\r{A}}. 2020, \apj, 900, 82

\bibitem[{{Rosen} {et~al.}(2020){Rosen}, {Offner}, {Sadavoy}, {Bhandare}, {V{\'a}zquez-Semadeni}, \& {Ginsburg}}]{Rosen2020}
{Rosen}, A.~L., {Offner}, S. S.~R., {Sadavoy}, S.~I., {et~al.} 2020, \ssr, 216, 62

\bibitem[{{Salii} {et~al.}(2022){Salii}, {Zinchenko}, {Liu}, {Sobolev}, {Aberfelds}, \& {Su}}]{Salii2022}
{Salii}, S.~V., {Zinchenko}, I.~I., {Liu}, S.-Y., {et~al.} 2022, \mnras, 512, 3215

\bibitem[{{Szymczak} {et~al.}(2018){Szymczak}, {Olech}, {Wolak}, {G{\'e}rard}, \& {Bartkiewicz}}]{Szymczak2018}
{Szymczak}, M., {Olech}, M., {Wolak}, P., {G{\'e}rard}, E., \& {Bartkiewicz}, A. 2018, \aap, 617, A80

\bibitem[{{Tan} {et~al.}(2014){Tan}, {Beltr{\'a}n}, {Caselli}, {Fontani}, {Fuente}, {Krumholz}, {McKee}, \& {Stolte}}]{Tan14}
{Tan}, J.~C., {Beltr{\'a}n}, M.~T., {Caselli}, P., {et~al.} 2014, Protostars and Planets VI, 149

\bibitem[{{Wang} {et~al.}(2011){Wang}, {Beuther}, {Bik}, {Vasyunina}, {Jiang}, {Puga}, {Linz}, {Rod{\'o}n}, {Henning}, \& {Tamura}}]{Wang11}
{Wang}, Y., {Beuther}, H., {Bik}, A., {et~al.} 2011, \aap, 527, A32

\bibitem[{{Wolf} {et~al.}(2024){Wolf}, {Stecklum}, {Caratti o Garatti}, {Boley}, {Fischer}, {Harries}, {Eisl{\"o}ffel}, {Linz}, {Ahmadi}, {Kobus}, {Haubois}, {Matter}, \& {Cruzalebes}}]{Wolf2024}
{Wolf}, V., {Stecklum}, B., {Caratti o Garatti}, A., {et~al.} 2024, \aap, 688, A8

\bibitem[{{Zinchenko} {et~al.}(2012){Zinchenko}, {Liu}, {Su}, {Kurtz}, {Ojha}, {Samal}, \& {Ghosh}}]{Zin12}
{Zinchenko}, I., {Liu}, S.-Y., {Su}, Y.-N., {et~al.} 2012, \apj, 755, 177

\bibitem[{{Zinchenko} {et~al.}(2015){Zinchenko}, {Liu}, {Su}, {Salii}, {Sobolev}, {Zemlyanukha}, {Beuther}, {Ojha}, {Samal}, \& {Wang}}]{Zin15}
{Zinchenko}, I., {Liu}, S.-Y., {Su}, Y.-N., {et~al.} 2015, \apj, 810, 10

\bibitem[{{Zinchenko} {et~al.}(2017){Zinchenko}, {Liu}, {Su}, \& {Sobolev}}]{Zin17}
{Zinchenko}, I., {Liu}, S.-Y., {Su}, Y.-N., \& {Sobolev}, A.~M. 2017, \aap, 606, L6

\bibitem[{{Zinchenko} {et~al.}(2020){Zinchenko}, {Liu}, {Su}, {Wang}, \& {Wang}}]{Zin2020}
{Zinchenko}, I.~I., {Liu}, S.-Y., {Su}, Y.-N., {Wang}, K.-S., \& {Wang}, Y. 2020, \apj, 889, 43

\end{thebibliography}

\end{document}